\documentclass[a4paper,12pt]{elsart}
\usepackage{epsfig}
\usepackage{subfigure}
\usepackage{amsmath}

\newcommand\ie{{\em i.e.}}

\newcommand\squark{{\tilde{q}}}
\newcommand\slepton{{\tilde{l}}}
\newcommand\selectronR{{\tilde{e}_R}}
\newcommand\ntlinoOne{{\tilde{\chi}_1^0}}
\newcommand\ntlinoTwo{{\tilde{\chi}_2^0}}

\newcommand\eref[1]{{(\ref{#1})}}

\begin{document}

\begin{frontmatter}

\title{Constrained invariant mass distributions in cascade decays.
  The shape of the ``$m_{qll}$-threshold'' and similar distributions.}

\author{Christopher G. Lester}
\address{Cavendish Laboratory, Cambridge}

\begin{abstract}
Considering the cascade decay $D\rightarrow c C \rightarrow c b B
\rightarrow c b a A$ in which $D,C,B,A$ are massive particles and
$c,b,a$ are massless particles, we determine for the shape of the
distribution of the invariant mass of the three massless particles
$m_{abc}$ for the sub-set of decays in which the invariant mass
$m_{ab}$ of the last two particles in the chain is (optionally)
constrained to lie inside an arbitrary interval, $m_{ab} \in [
m_{ab}^\text{cut min}, m_{ab}^\text{cut max}]$.  An example of an
experimentally important distribution of this kind is the ``$m_{qll}$
threshold'' -- which is the distribution of the combined invariant
mass of the visible standard model particles radiated from the
hypothesised decay of a squark to the lightest neutralino via
successive two body decay,: $\squark \rightarrow q \ntlinoTwo
\rightarrow q l \slepton \rightarrow q l l \ntlinoOne $, in which the
experimenter requires additionally that $m_{ll}$ be greater than
${m_{ll}^{max}}/\sqrt{2}$.  The location of the ``foot'' of this
distribution is often used to constrain sparticle mass scales.  The
new results presented here permit the location of this foot to be
better understood as the shape of the distribution is derived.  The
effects of varying the position of the $m_{ll}$ cut(s) may now be seen
more easily.
\end{abstract}

\end{frontmatter}



\def\slashchar#1{\setbox0=\hbox{$#1$}           
   \dimen0=\wd0                                 
   \setbox1=\hbox{/} \dimen1=\wd1               
   \ifdim\dimen0>\dimen1                        
      \rlap{\hbox to \dimen0{\hfil/\hfil}}#1 
   \else                                        
      \rlap{\hbox to \dimen1{\hfil$#1$\hfil}}/                                    \fi}          
\newcommand\ptmiss{\slashchar{p}_T}
\newcommand\Ptmiss{\slashchar{{\bf p}}_T}
\newcommand\pmiss{{{\slashchar{p}}}}
\newcommand\mabc{{m_{abc}}}
\newcommand\mabcsq{{m_{abc}^2}}
\newcommand\mab{{m_{ab}}}
\newcommand\mabsq{{m_{ab}^2}}
\newcommand\mac{{m_{ac}}}
\newcommand\macsq{{m_{ac}^2}}
\newcommand\mbc{{m_{bc}}}
\newcommand\mbcsq{{m_{bc}^2}}
\newcommand\mabmax{{m_{ab}^\text{max}}}
\newcommand\mabcmax{{m_{abc}^\text{max}}}
\newcommand\mabmaxsq{{(m_{ab}^\text{max})^2}}
\newcommand\mabcmaxsq{{(m_{abc}^\text{max})^2}}
\newcommand\mabcthresh{{m_{abc}^\text{thresh}(x)}}
\newcommand\mabcthreshsq{{(m_{abc}^\text{thresh}(x))^2}}
\newcommand\dgammadmabcsq{{\frac{d \Gamma}{d(\mabcsq)}}}
\newcommand\dgammadmabc{{\frac{d \Gamma}{d(\mabc)}}}
\newcommand\pmabcsq{{P(\mabcsq)}}
\newcommand\mqlhigh{{m_{ql}^\text{high}}}
\newcommand\mqllow{{m_{ql}^\text{low}}}
\newcommand\mqlhighsq{{(\mqlhigh)^2}}
\newcommand\mqllowsq{{(\mqllow)^2}}
\newcommand\ds{{m_D^2}}
\newcommand\cs{{m_C^2}}
\newcommand\bs{{m_B^2}}
\newcommand\as{{m_A^2}}
\newcommand\md{{m_D}}
\newcommand\mc{{m_C}}
\newcommand\mb{{m_B}}
\newcommand\ma{{m_A}}
\newcommand\maxcut{{m_{ab}^\text{cut max}}}
\newcommand\mincut{{m_{ab}^\text{cut min}}}
\newcommand\maxcutsq{{(\maxcut)^2}}
\newcommand\mincutsq{{(\mincut)^2}}
\newcommand\maxboundsq{{(m_{ab}^{+})^2}}
\newcommand\minboundsq{{(m_{ab}^{-})^2}}
\newcommand\maxminboundsq{{(m_{ab}^{\pm})^2}}
\newcommand\nlep{{n_{\rm leptons}}}
\def\ptlepOne{{p_T^{l_1}}}
\def\ptlepTwo{{p_T^{l_2}}}
\def\PtlepOne{{{\bf p}_T^{l_1}}}
\def\PtlepTwo{{{\bf p}_T^{l_2}}}
\def\ptjetI{{p_T^{j_i}}}
\def\PtjetI{{{\bf p}_T^{j_i}}}
\def\ptjetOne{{p_T^{j_1}}}
\def\ptjetTwo{{p_T^{j_2}}}
\def\ptjetThree{{p_T^{j_3}}}
\def\ptjetFour{{p_T^{j_4}}}
\def\GeV{{{\rm {GeV}}}}
\def\njet{{n_{\rm jets}}}
\def\etmiss{\slashchar{E}_T}

\section{Introduction}

\begin{figure}
\centerline{
\subfigure[]{\psfig{file=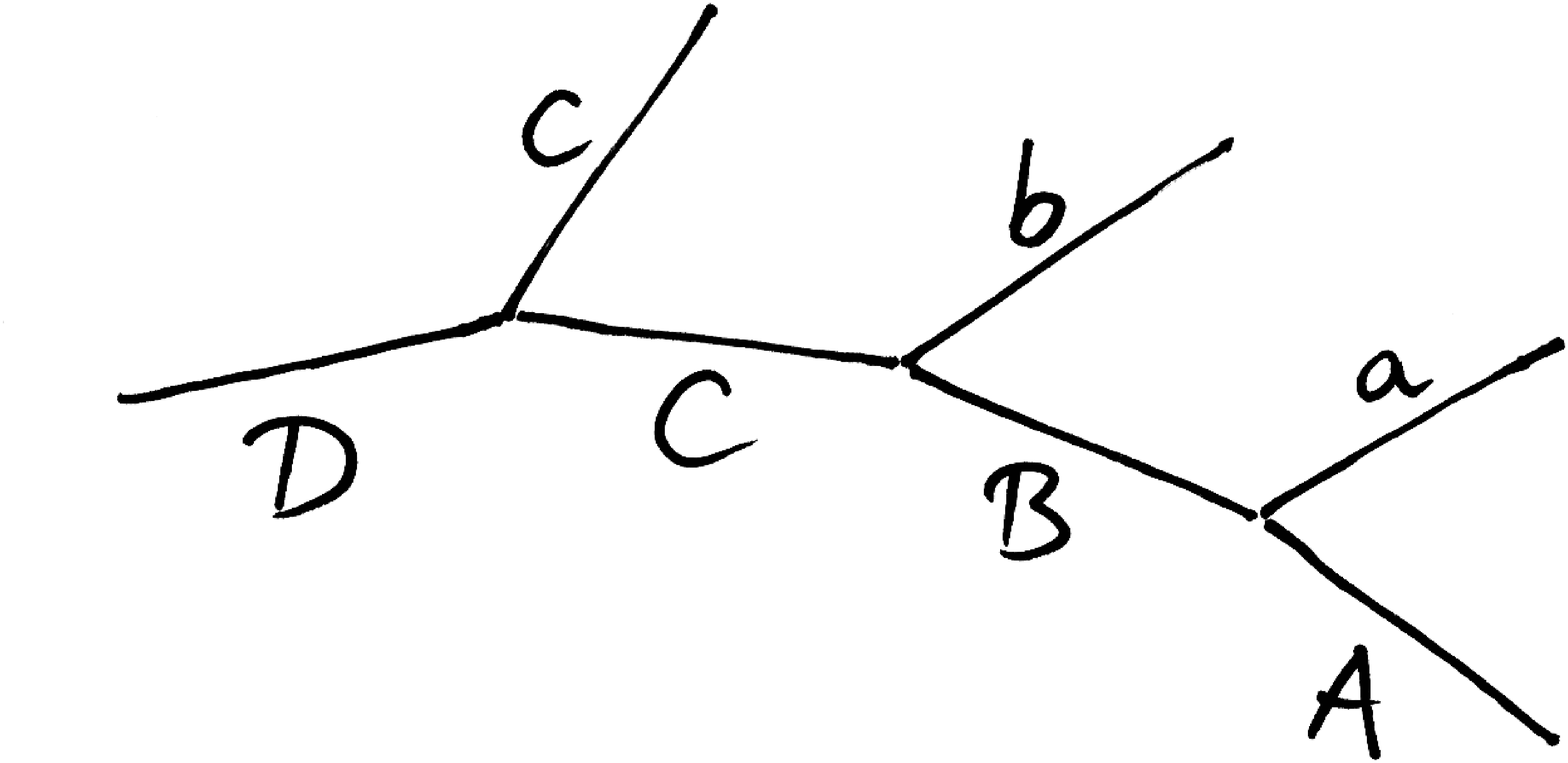,height=4cm}}
\subfigure[]{\psfig{file=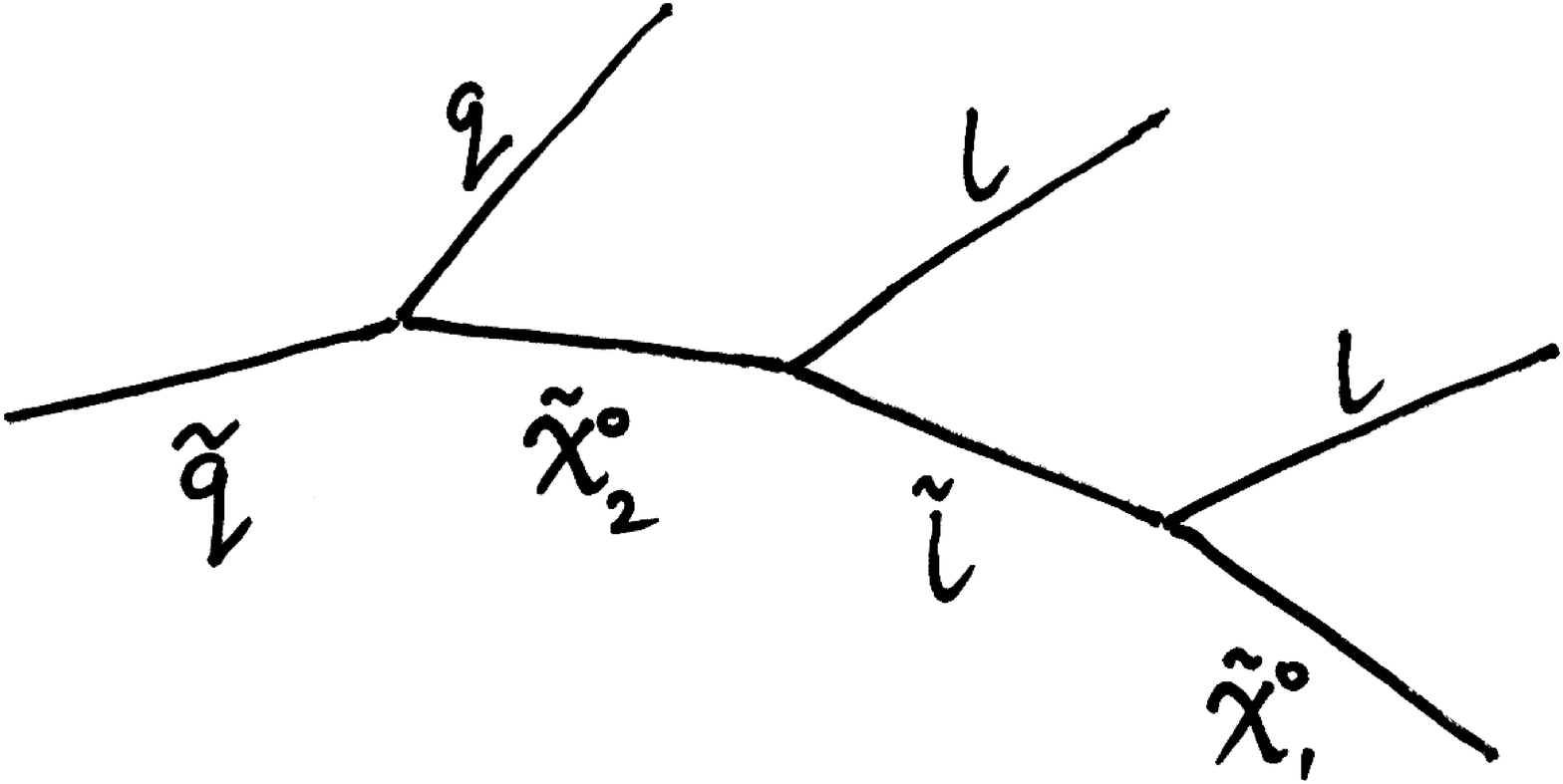,height=4cm}}
}
\caption{The decay chain under study.  Part (a) shows the chain in
  ``generic'' form with the massive particles on the ``backbone''
  labelled with capital letters, and the massless visible particles
  labelled in lower case.
Part (b) shows a commonly hypothesised squark decay chain open in many
  supersymmetric models.  This particular chain is the example most
  frequently used to motivate the study of the ``generic'' chain in
  part (a).}
\label{fig:decayChain}
\end{figure}

We consider the cascade decay $D\rightarrow c C \rightarrow c b B
\rightarrow c b a A$ shown in figure~\ref{fig:decayChain} in which
$D,C,B,A$ are massive particles and $c,b,a$ are treated as massless
particles.

In the context of the Large Hadron Collider (LHC), the most studied example
of such a decay chain is the hypothesised decay of a squark to the
lightest neutralino via three successive two body decays: $\squark
\rightarrow q \ntlinoTwo \rightarrow q l \slepton \rightarrow q l l
\ntlinoOne $.  The decay chain is relevant to more than just
supersymmetric decays, however.  For example it may appear in models
with extra dimensions, in which the particles would be Kaluza-Klein
excitations.

It has been frequently suggested
\cite{Bachacou:1999zb,AtlasTDR,Lester,Allanach:2000kt,Lester:2005je} that one of the main ways we might
hope to measure or constrain the masses of new particles produced at
the LHC will be through the study of kinematic endpoints of invariant
masses constructed from the visible products of decay chains like
those in figure~\ref{fig:decayChain}.

Given the observed four-momenta $a^\mu$, $b^\mu$ and $c^\mu$ it is in
principle possible to construct three non-trivial linearly-independent
Lorentz-invariant quantities, namely $m_{ab}^2 = (a^\mu+b^\mu)^2$,
$m_{bc}^2 = (b^\mu+c^\mu)^2$ and $m_{ac}^2 = (a^\mu+c^\mu)^2$.  A
fourth quantity $m_{abc}^2 = (a^\mu + b^\mu + c^\mu)^2$ is related to
the first three via $m_{abc}^2 = m_{ab}^2+m_{bc}^2+m_{ac}^2$.

In many models in which this decay chain is present, however, it is
not possible to tell two of the particles apart (typically $a$ and
$b$).  For example, in the supersymmetric decay chain $\squark
\rightarrow q \ntlinoTwo \rightarrow q l \slepton \rightarrow q l l
\ntlinoOne $ the particles $a$ and $b$ form a lepton-antilepton pair,
and it may be impossible to say which observed lepton was the ``$a$''
and which was the ``$b$''.\footnote{If it is possible to measure the
charge of the quark-jet some small fraction of the time, then there is
a small discriminatory power in the $a$/$b$ assignment to the observed
leptons, if the spins of the sparticles are known.  See
\cite{Barr:2004ze} and subsequent papers.}  In such models, the three
independent variables are instead $\mabsq$, $\mqlhighsq =
\max{[\macsq, \mbcsq]}$ and $\mqllowsq = \min{[\macsq, \mbcsq]}$.  We
see that the experimentally unidentifiable $\macsq$ and $\mbcsq$ have
been replaced by experimentally observable $\mqlhighsq$ and
$\mqllowsq$.  Note that as before, $m_{abc}^2 =
m_{ab}^2+\mqlhighsq+\mqllowsq$.

Although there are only three independent variables that categorise a
given decay, by looking at {\em more than one event of the given decay
type} it is often possible to construct {\em more than three}
independent constraints on the masses of the particles involved.
This has been done most frequently by constructing new variables
which are arbitrary functions of the three independent variables, and
then looking for the kinematic endpoints\footnote{i.e.\ the largest or
smallest possible values that such a variable can take in a physically
allowed decay configuration} of the distributions of these variables
over a large number of events.  

In principle the {\em shape} of any one of these distributions
contains more information than the position of the associated
kinematic endpoint.  Nonetheless, in previous years there has been a
tendency to concentrate mainly on constraints coming from kinematic
endpoints because interpreting these constraints does not require a
detailed understanding of the way detector acceptance and
reconstruction efficiency may vary {\em across} any of these
distributions.  Such detector effects may not be known with
great confidence until a number of years after LHC turn-on.

Recently, however, there has been a resurgence of interest in
understanding the shapes of these kinematic distributions
\cite{Gjelsten:2004ki,Miller:2005zp} not least because an
understanding of the shapes of these distributions is a pre-requisite
for being able to fit the positions of the kinematic endpoints, even
if one chooses to ignore all the information contained in the rest
of the shape.  In particular \cite{Miller:2005zp} has recently
calculated the shapes of the distributions of the $m_{ab}$,
$\mqlhigh$, $\mqllow$ and $\mabc$ distributions.  One distribution
whose shape was not calculated in \cite{Miller:2005zp} or elsewhere,
however, was the shape of the $\mabcthresh$ distribution.

Specifically the $\mabcthresh$ distribution is a constrained form of
the $\mabc$ distribution in which the only events which are considered
are those for which the value of $m_{ab}$ happens to exceed a cut $x =
\mincut$ which is most frequently chosen to be $1/\sqrt 2$ times the
maximum value which $\mab$ can take.\footnote{There is a long
outstanding question (see remark in \cite{Lester}) as to whether this
somewhat arbitrary choice for $\mincut$ is indeed the one that gives
the best constraint on the masses of the particles involved in the
decays.}  The {\em lower} kinematic endpoint is the one whose position
is most sensitive to the additional constraint (although the position
of the upper endpoint can also be modified), and hence it is called a
``threshold'' rather than a ``maximum''.  The $\mabcthresh$
distribution is used because the position of the threshold, as a
function of the masses of the particles involved, is more non-linear
than the locations of other endpoints.  This means that the position
of the threshold can sometimes provide a handle on the absolute mass
scale of the particles involved.  Without a constraint from the
endpoint of the $\mabcthresh$ distribution, the other edges usually
only succeed in constraining mass differences.

Though important, the shape of threshold of the $\mabcthresh$
distribution has been poorly studied in the past.  Poor understanding
of the shape of this edge has frequently resulted in the location of
this endpoint being assigned a large systematic error (see for example
\cite{Bachacou:1999zb,AtlasTDR,Lester,Allanach:2000kt,Lester:2005je}).
Even allowing for a larger systematic error on the edge measurement,
it often seems to be the case that the displacement of this edge from
its ``nominal'' value is greater than a naive interpretation of the
edge as a straight line would lead one to expect.

We redress the historical imbalance in this note, by determining the
shape of the $\mabcthresh$ distribution in the spinless approximation,
which is valid for supersymmetry at the LHC.  We do this for arbitrary
values of $x=\mincut$, the imposed lower-bound requirement on $\mab$.
Given the form of the method, it turns out to be no extra effort to
allow also an arbitrary upper-bound requirement $\maxcut$ on $\mab$,
should one wish to exploit the information contained in the changes to
the upper part of the distribution caused by varying this quantity.

\subsection{Existing results from earlier work}

First we note some existing results which we will make use of in later
sections.  \cite{AtlasTDR} (and many of the references in the
Supersymmetry Chapter therein) told us that:
\begin{equation}
\mabmaxsq  =  \frac{(\cs-\bs)(\bs-\as)}{\bs},
\end{equation}
Also \cite{Lester,Allanach:2000kt} told us that
\begin{equation}
\mabcmaxsq  = \begin{cases}
(\ds-\cs)(\cs-\as)/\cs   &  \text{iff $\cs < \ma \md$}, \\
(\ds-\bs)(\bs-\as)/\bs   &  \text{iff $\ma \md < \bs$}, \\
(\ds \bs - \cs \as)(\cs - \bs)/(\cs \bs) &  \text{iff $\bs \md < \ma \cs$}, \\
(\md - \ma)^2    &  \text{otherwise.}
\end{cases}
\end{equation}
It was also shown in \cite{Lester} and more recently in \cite{Kitano:2006gv} that
\begin{equation}
\mabcthreshsq =  x^2 + 2 E^{(C)}_c (E_{ab}^{(C)}(x)-p_{ab}^{(C)}(x))
\end{equation}
where
\begin{eqnarray}
E_c^{(C)} & = & \frac{\ds-\cs}{2\mc}, \label{eq:ecC}\\
(p_{ab}^{(C)}(x))^2 & = & \frac{(\cs+\as-x^2)^2-4 \cs \as}{4 \cs}\qquad\quad\text{and}\label{eq:pabC}\\
E_{ab}^{(C)(x)} & = & \sqrt{ x^2 + (p_{ab}^{(C)}(x))^2} \label{eq:eabC}\\ 
& = & \frac{x^2 + (\cs-\as)}{2 \mc},
\end{eqnarray}
where $x=\mincut$ is the smallest value of $\mab$ which events are
allowed to have if they are to become part of the ``threshold distribution''.

\section{Result}

The main result of this paper is that:
\begin{equation}
\frac{1}{\Gamma_{\text{Tot}}}\dgammadmabcsq = \pmabcsq
\end{equation}
where
\begin{equation}
\pmabcsq =  
\begin{cases}
\left[ {
J(\beta)-J(\alpha)
} \right]
\Theta\left({\beta-\alpha}\right)/Z \label{eq:mainresult} &
\text{if $\mabcsq \le (\md-\ma)^2$} \\
0 & \text{otherwise.} \\
\end{cases}
\end{equation}
In the above we have defined the function
\begin{equation}
J(x) = - \cosh^{-1} \left(\frac{\cs+\as-x}{2\mc\ma}\right)
\end{equation}
and defined the quantities
\begin{eqnarray}
\beta  & = & \min \left[  \maxcutsq, \maxboundsq, \mabmaxsq \right],\label{eq:beta}\\
\alpha & = & \max \left[  \mincutsq, \minboundsq, 0 \right]\label{eq:alpha}
\end{eqnarray}
in which
\begin{equation}
\maxminboundsq = \frac{1}{2 \ds} \left[
{
(\ds+\cs)\mabcsq
-
(\ds-\cs)\left\{
(\ds-\as) \mp \sqrt{X^4}\label{eq:hasroot}
\right\}
}
\right]
\end{equation}
in which
\begin{equation}
X^4 = (\ds + \as - \mabcsq)^2-4 \as \ds. \label{eq:X}
\end{equation}

If the normalising constant $Z$ in \eref{eq:mainresult}\ is chosen to
be
\begin{equation}
Z = \left(\frac{\ds-\cs}{\cs}\right)\left(\maxcutsq-\mincutsq\right),\label{eq:normconst}
\end{equation}
then the integral of $\pmabcsq$ over physical values of $\mabcsq$ is
fixed at unity.  This makes $\pmabcsq$ into a {\em normalised
probability distribution}, suitable for use in log likelihood fits,
etc.  The function $\Theta(x)$ in \eref{eq:mainresult}\ is the
Heaviside Step Function, equal to 0 for $x<0$ and equal to 1 for
$x>=0$.  If desired, the normalised distribution in mass space (rather
than mass-squared space) is easily found via
\begin{equation}
\frac{1}{\Gamma_{\text{Tot}}}\dgammadmabc = 2 \mabc
\frac{1}{\Gamma_{\text{Tot}}} \dgammadmabcsq.
\end{equation}
\subsection*{Remarks}

For the ``classic'' $m_{qll}$-threshold of \cite{Bachacou:1999zb,AtlasTDR,Lester,Allanach:2000kt,Lester:2005je}, the following choices are made:
\begin{eqnarray}
\maxcutsq & = & \mabmaxsq,\\
\mincutsq & = & \mabmaxsq/2. \label{eq:changeme}
\end{eqnarray}

One can recover the standard ``full'' $\mabcsq$ distribution (\ie\ the
distribution whose {\em upper} edge provides the measurement, not the
constrained threshold distribution derived from it by applying extra
cuts) by setting $\mincutsq$ in \eref{eq:changeme} to zero.

\par

Neither the final argument to the $\min$ in \eref{eq:beta}\ nor the
final argument to the $\max$ in \eref{eq:alpha}\ is strictly
necessary.  These arguments are there only to protect against
accidental unphysical choices of $\maxcutsq$ and $\mincutsq$, \ie\
values which do not satisfy
\begin{equation}
0 \le \mincutsq \le \maxcutsq \le
\mabmaxsq.\label{eq:physicality}
\end{equation}
The normalising constant $Z$ in \eref{eq:normconst} has not been
similarly protected, so care should be taken to ensure that
\eref{eq:physicality} is satisfied when evaluating $Z$.

If the function $J(x)$ is only evaluated when \eref{eq:mainresult}\
demands it, $X^4$ will always be non-negative and its square root may
safely be taken in \eref{eq:hasroot}.

Despite what the notation may appear to imply, either (or indeed both)
of $\maxboundsq$ and $\minboundsq$ can be negative.  It is important
to retain the signs of these quantities when they are used in equations
\eref{eq:beta}\ and \eref{eq:alpha}.

\begin{figure}
\centerline{
\psfig{file=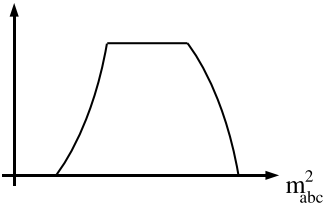,height=3cm}}
\caption{Cartoon of the shape of the $\mabcsq$ distribution defined in
    \eref{eq:mainresult}. Note the distribution is over {\em the
    square of the mass}, not over the mass itself.}
\label{fig:simplified}
\end{figure}

In mass-squared-space (\ie\ plotting $\dgammadmabcsq$ rather than
$\dgammadmabc$) the shape of the distribution resembles a ``top-hat
distribution with sloping sides'' as shown in
figure~\ref{fig:simplified}.  The flat upper part of the top-hat begins
and ends at roots of $\maxcutsq = \maxboundsq$ or $\mincutsq =
\minboundsq$ which are the places at which the $max$ and $min$
functions in \eref{eq:beta} and \eref{eq:alpha} have cusps.  Though
not straight lines, the ``flanks'' of the distribution are often not
strongly curved, and so in the absence of effects from resolution,
detector acceptance and other cuts, and in the absence of contamination
from non-pure chains the lower edge should vanish linearly.  This
effect is not altered significantly by the transition from
mass-squared space to mass space, unless the threshold is itself very close
to the origin, whereupon the edge takes on a quadratic character.

\section{Selection cuts and impurities}

Event selection cuts (and other effects) modify the
shape of the $m_{qll}$-threshold distribution.

We illustrate this by generating a 100 inverse femtobarn sample of
general SUSY events using Herwig 6.1 \cite{Corcella:2000bw} and pass
them through a simple detector simulation \cite{atlfast} which crudely
models the response of the ATLAS LHC experiment.  Following
\cite{Bachacou:1999zb,AtlasTDR,Lester,Allanach:2000kt} and others, we
work with the minimal SUGRA model known as ``SUGRA Point 5'' defined
by $m_0=100~\GeV$, $m_{\frac 1 2}=300~\GeV$, $A_0 = 300~\GeV$, $\tan
\beta = 2.1$ and $\mu > 0$.  This model has $m_\ntlinoOne = 121.5~\GeV$, $m_\selectronR = 157.2~\GeV$, $m_\ntlinoTwo= 233~\GeV$ and
has squarks of order $650-660~\GeV$ (although the lightest stop is
much lighter at $494~\GeV$).

When constructing the ``experimentally observed'' $m_{qll}$-threshold
distribution, we use exactly the procedure described in \cite{Lester},
which is to say in summary:
\begin{itemize}
\item We require $\nlep = 2$, both leptons are opposite-sign
  same-family (OSSF) and $\ptlepOne \ge \ptlepTwo \ge 10~\GeV$.
\item
We require $\njet \ge 2$ and $\ptjetOne \ge \ptjetTwo \ge 150~\GeV$.
\item We require $\ptmiss \ge 300\ \GeV$.
\item
We form only the {\em larger} of the two possible $m_{qll}$ combinations
(one for each of $j_1$ and $j_2$) as our main interest is in
preserving the quality of the lower edge, which we do not wish to
obscure with wrong-combination backgrounds.  This jet-choice ensures
wrong-choice backgrounds fall at high rather than at low mass values.
\item
We repeat the above set of selection cuts but for opposite-sign
different-family (OSDF) leptons.  We add these events with weight -1 to
the final histogram in order to subtract backgrounds from uncorrelated
leptons.
\end{itemize}

\begin{figure}
\centerline{
\subfigure[]{\psfig{file=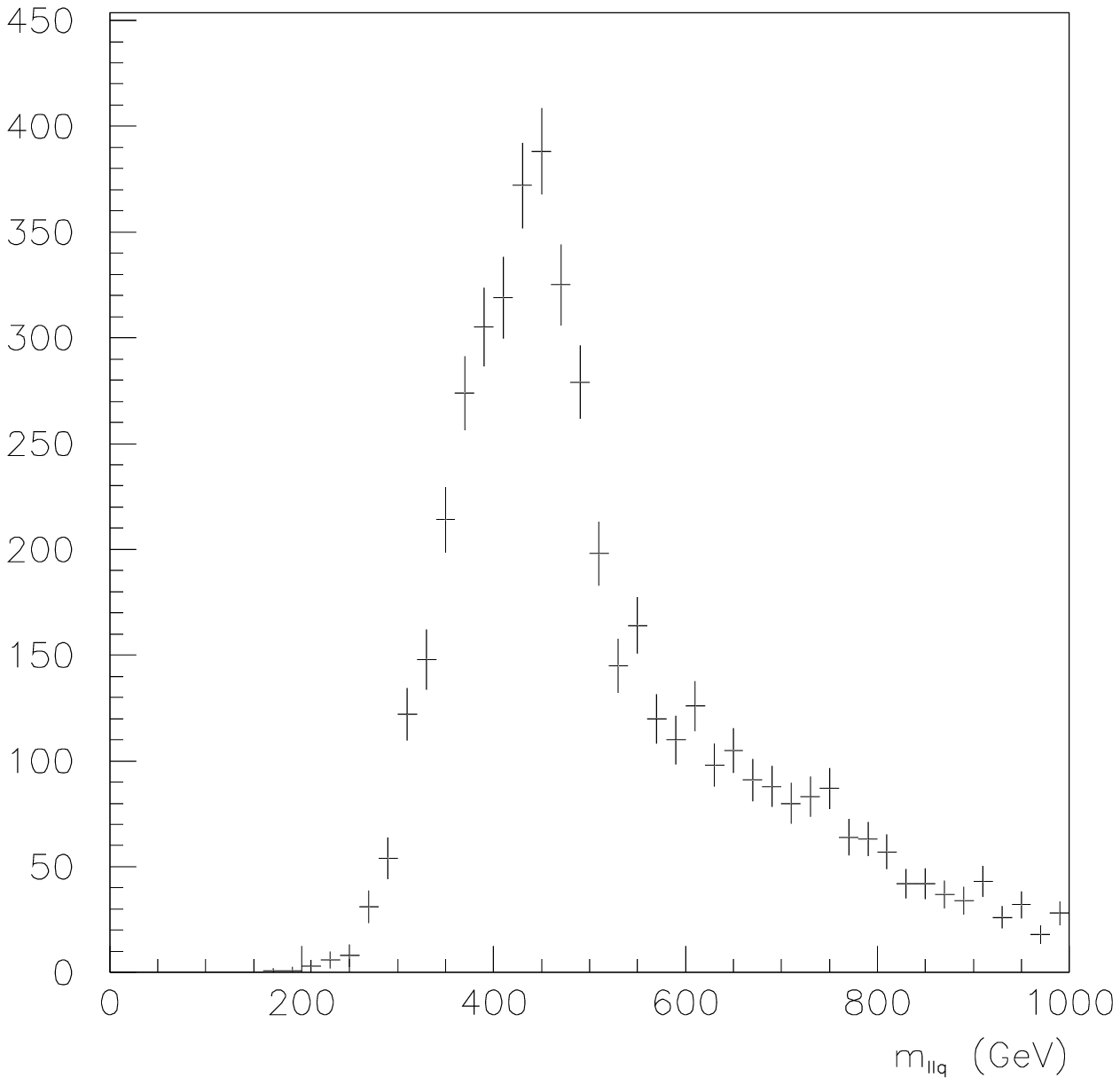,height=5cm}}
\subfigure[]{\psfig{file=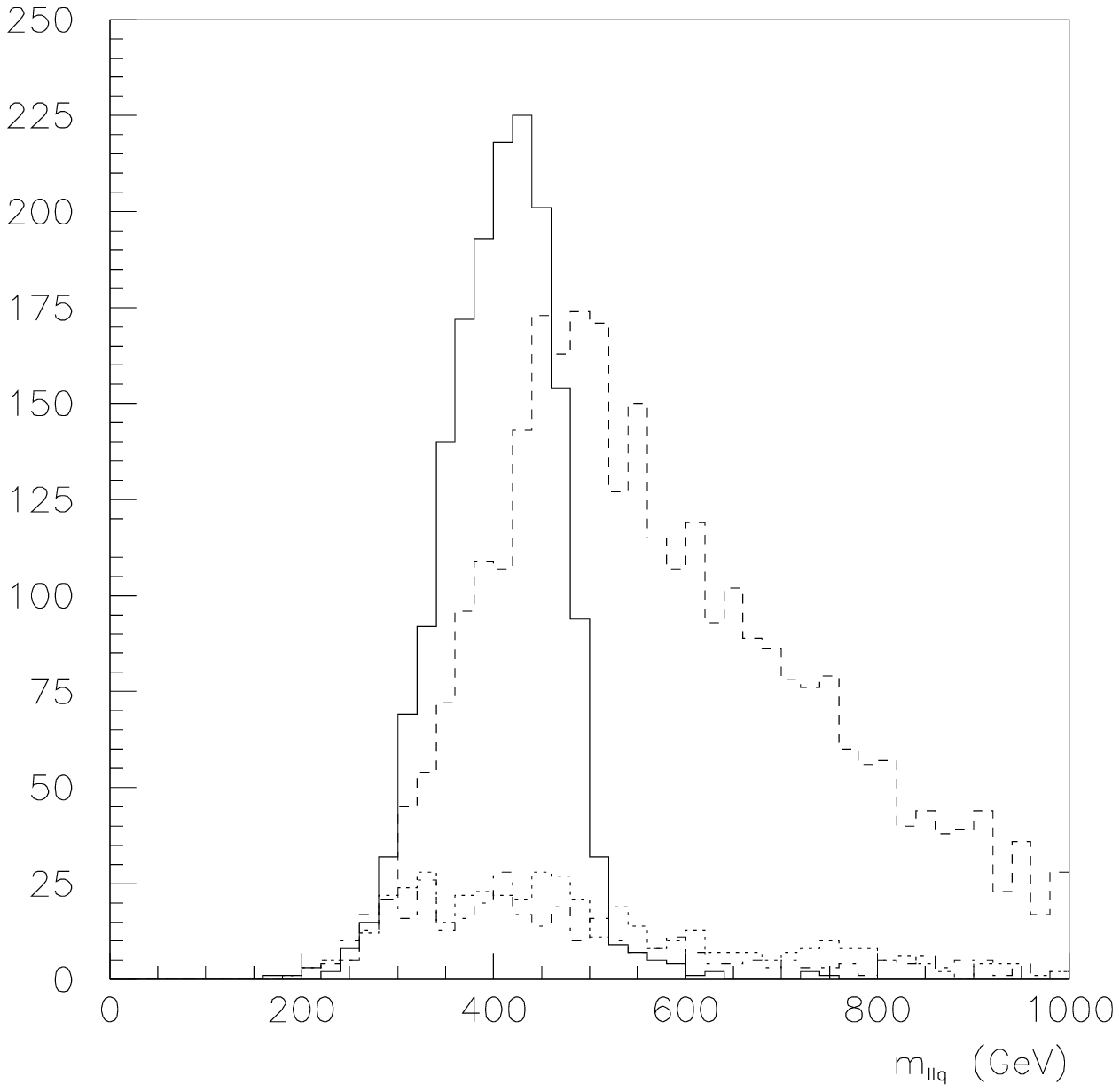,height=5cm}}
\subfigure[]{\psfig{file=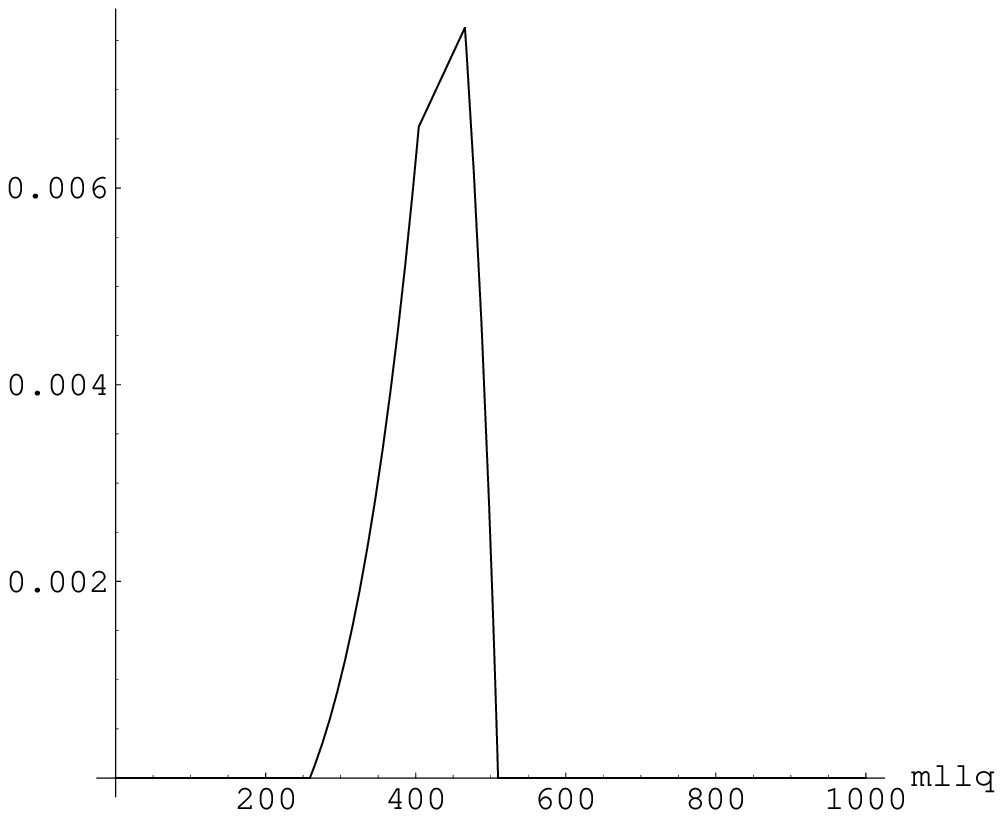,height=4.6cm}}
}
\caption{Three sets of $m_{qll}$-threshold distributions for the CMSSM
mSUGRA model described in the text.  (a) shows what might be seen in a
typical LHC experiment after the cuts described in the text.  Figure
(b) shows (a) broken down into the four components which contribute to
it.  These are (1) a narrow component coming from events in which the
jet was {\em correctly} identified (solid line), (2) a broad component
in which the jet was {\em incorrectly} identified (dashed line), (3) a
small component representing the contribution from all SUSY sources
which have nothing to do with the desired chain (dotted line), and
finally (4) a small component from all SUSY sources which pass
opposite-sign {\em different-family} versions of the cuts (dot-dashed
line).  The shape of component (4) matches the shape of (3) nicely and
is added with negative weight in (a) to subtract component (3).
Figure (c) shows the shape of the $m_{qll}$-threshold distribution as
predicted in the text using the masses $m_\ntlinoOne = 121.5$,
$m_\slepton = 157.2$, $m_\ntlinoTwo= 233$ and $m_\squark =
656~\GeV$. }
\label{fig:breakdown}
\end{figure}

The resultant ``experimentally observed'' $m_{qll}$-threshold
distribution may be seen in figure~\ref{fig:breakdown}(a).
Figure~\ref{fig:breakdown}(b) breaks this distribution down into the
components in which the jet was correctly or incorrectly identified,
and figure~\ref{fig:breakdown}(c) shows the distribution predicted by
the mass-space form of \eref{eq:mainresult} using the masses
$m_\ntlinoOne = 121.5$, $m_\slepton = 157.2$, $m_\ntlinoTwo= 233$ and
$m_\squark = 656~\GeV$.  The first three masses are taken directly
from the model.  The final mass represents the mass of a ``typical''
squark - an average of the masses of the twelve squarks participating
in the chain.

The figure shows that there is good agreement between the analytic
shape in \ref{fig:breakdown}(c) and the ``correct combination''
component of the ``experimentally observed'' distribution, even though
the jet-selection method should introduce a slight skew to the
``correct combination'' component.  It seems that for these masses the
skew is not very noticeable.  The presence of the significant ``wrong
combination'' background, however, means that it is important not to
use the analytic shape on its own.

When the $m_{qll}$-threshold analysis was first proposed
\cite{Bachacou:1999zb} the analytic form of the shape was not known.
This motivated the deliberately high-mass-biasing jet-choice which
seeks to keep the threshold of the mass distribution clean.  Now that
the shape of the distribution is known, it may prove more fruitful in
later studies to bin {\em both} jet combinations, and then fit the
analytic shape on top of a continuum background whose shape could
perhaps be estimated from the side bands.

\clearpage

\section{Conclusions}

The shape of the distribution of the total invariant mass of the light
particles emitted in the decays shown in figure~\ref{fig:decayChain}
has been calculated for the sub-set of decays which pass an additional
upper and/or lower cut on the invariant mass of the last two particles
emitted in the decay.  This distribution is known generically as the
$m_{abc}$-threshold distribution, or in its most commonly used
specific example as the $m_{qll}$-threshold distribution.  Even though
the calculation treats all particles involved as scalars, the result
is still {\em exact} in many cases, and will remain a good
approximation in many others.  The ``classic'' $m_{qll}$-threshold
distribution in supersymmetry is one of the places where the spinless
approximation is {\em exact}, as spin effects from $ql^+l^-$ cancel
those from $ql^-l^+$.  Deviations from the spinless approximation are
potentially observable in UED and other non-susy models where spin
effects in $q^*l^+l^-$ and $q^*l^-l^+$ do not cancel completely,
although the magnitude of these deviations are likely to be small as
the production excess of $q^*$ over ${\bar q}^*$ is itself expected to
be small at the LHC over much of parameter space.

\section{Acknowledgements}

The author would like to thank D.\ T.\ Koch, A.\ R.\ Raklev and
members of the Cambridge Supersymmetry Working Group (in particluar
J.\ Smillie and A.\ J.\ Barr) for helpful comments and discussions.

\section{Appendix}

The derivation of the result \eref{eq:mainresult}\ may be achieved
using the following steps.
\begin{itemize}
\item
Work in the rest frame of the $C$.
\item
In this frame the energy of the $c$ is calculable (see \eref{eq:ecC}),
as is the energy of the $(a+b)$-system if it is taken to be an
effective particle of known mass $m_{ab}$ (see \eref{eq:eabC}).
\item
The mass-squared $m_{ab}^2$ may easily be shown to be distributed
uniformly, and the polar angle $\theta$ between the $c$ and the
$(a+b)$-system in this frame will be uniform in $\cos\theta$ if either
the spin of the $C$ may be neglected or if we are unable (or choose
not) to distinguish contributions from different spins.  This will happen
at the LHC where no distinction will be drawn between $e^+ e^-$ events
and $e^- e^+$ events.  As a result we have:
\begin{equation}
\Gamma \propto \int_I d{(\cos\theta)}d{(\mabsq)}.
\end{equation}
\item
Change variables from $\cos\theta$ to $\mabcsq$, resulting in
\begin{equation}
\Gamma \propto \int_I \frac{1}{2 \mc p_{ab}^{(C)}} d{(\mabcsq)}d{(\mabsq)}.
\end{equation}
\item
The boundary of the region $I$ may be shown in
$(\mabsq,\mabcsq)$-space to be the intersection of
the interior of one half of a hyperbola with a vertical strip of
constant width in $\mabsq$ (starting at $\mincutsq$ and ending at
$\maxcutsq$).  
\item
It is possible to write the integration limits for $I$ as upper and
lower bounds on $\mabsq$ (respectively $\beta$ and $\alpha$ in
\eref{eq:beta}\ and \eref{eq:alpha}) for given $\mabcsq$, allowing the
order of integration to be exchanged, and the integral over $\mabsq$
to be performed analytically:
\begin{eqnarray}
\Gamma & \propto & \int \int_\alpha^\beta
\frac{1}{2 \mc p_{ab}^{(C)}} d{(\mabsq)}d{(\mabcsq)} \\
& = & \int \left[-cosh^{-1}\left(\frac{\cs + \as - x}{2 \mc \ma}\right)\right]^{x=\beta}_{x=\alpha} d{(\mabcsq)}.
\end{eqnarray}
from which the form of $\pmabcsq$ can be read directly.
\item
The value of $Z$ needed to normalise $\pmabcsq$ to unit area is found
by performing the remaining integral.
\end{itemize}

\bibliography{draft}

\end{document}